\documentstyle[12pt]{article}
\newcommand\lsim{\mathrel{\rlap{\lower4pt\hbox{\hskip1pt$\sim$}}
    \raise1pt\hbox{$<$}}}
\newcommand\gsim{\mathrel{\rlap{\lower4pt\hbox{\hskip1pt$\sim$}}
    \raise1pt\hbox{$>$}}}

\begin{document}
\title{A General Analytic Formula for the \\ Spectral Index of the Density
Perturbations produced during Inflation}
\author{Misao Sasaki \\ Department of Earth and Space Science \\
Osaka University \\ Toyonaka 560 \\ Japan \and
Ewan D. Stewart\thanks{Present address:
Research Center for the Early Universe, School of Science,
University of Tokyo, Bunkyo-ku, Tokyo 113, Japan.
Email:~eds@utaphp1.phys.s.u-tokyo.ac.jp} \\
School of Physics and Chemistry \\ Lancaster University \\
Lancaster, LA1 4YB \\ U.K.}
\maketitle
\begin{abstract}
The standard calculation of the spectrum of density perturbations produced
during inflation assumes that there is only one real dynamical degree of
freedom during inflation.
However, there is no reason to believe that this is actually the case.
In this paper we derive general analytic formulae for the spectrum and
spectral index of the density perturbations produced during inflation.
\end{abstract}
\vspace*{-105ex}
\hspace*{\fill}{LANCASTER-TH/9504}\\
\hspace*{\fill}{OU-TAP-22}\hspace*{5.3em}\\
\hspace*{\fill}{astro-ph/9507001}\hspace*{2.3em}
\thispagestyle{empty}
\setcounter{page}{0}
\newpage
\setcounter{page}{1}

\section{Introduction}

During inflation \cite{inflation} vacuum fluctuations on scales less than the
Hubble radius in scalar fields with effective masses much less than the Hubble
parameter\footnote{All scalar fields generically acquire effective masses at
least of the order of the Hubble parameter in the early Universe
\cite{mass,fvi,Dine}. However, this can be naturally avoided for some scalar
fields during inflation in certain classes of supergravity theories
\cite{iss,Murayama}.} are magnified into classical perturbations in the scalar
fields on scales larger than the Hubble radius.
These classical perturbations in the scalar fields can then change the
number of $e$-folds of expansion and so lead to classical
curvature/density perturbations after inflation.
These density perturbations are thought to be responsible for the formation of
galaxies and the large scale structure of the observable Universe as well as,
in combination with the gravitational waves produced during inflation, for the
anisotropies in the cosmic microwave background.

The standard calculation \cite{single,index,Mukhanov,2nd,dhlrep} of the
spectrum of density perturbations produced during inflation assumes that there
is only one real dynamical degree of freedom during inflation.
Although this is the case in most of the models of inflation constructed up
to now, it is by assumption rather than prediction.
When one tries to construct models of inflation \cite{fvi,iss,mhi} that might
arise naturally in realistic models of particle physics, such as the low
energy effective supergravity theories derived from superstrings, one often
gets more than one dynamical degree of freedom during inflation.
The standard calculation is then generally not applicable.

In this paper we derive general analytic formulae for the spectrum and
spectral index of the density perturbations produced during inflation.
This work is based on earlier work by Starobinsky \cite{Star}.
See also \cite{other}.
While this work was in slow preparation three other related papers
\cite{Yokoyama,Polarski,Wands} appeared in the archives.
After this work was completed another related paper \cite{Salopek} appeared
in the astro-ph archive.

\section{Gravity}

We assume the gravitational part of the action to be
\begin{equation}
S = - \frac{1}{2} \int R \sqrt{-g}\, d^4x \,.
\end{equation}
It seems unlikely that non-Einstein gravity is relevant to inflation
because the inflation that inflated the observable Universe beyond the Hubble
radius must have occurred at an energy scale well below the Planck scale.
We therefore do not consider it.
We set $ M_{\rm Pl} = 1 / \sqrt{8\pi G} = 1 $ throughout the paper.

\subsection{The background}

The background metric is
\begin{equation}
ds^2 = dt^2 - a(t)^2 \delta_{ij} dx^i dx^j \,.
\end{equation}
Two important quantities are the Hubble parameter
\begin{equation}
H = \frac{\dot{a}}{a}
\end{equation}
and the number of $e$-folds of expansion
\begin{equation}
N = \int H dt \,.
\end{equation}

\subsection{${\cal R}$}

Scalar linear perturbations to the metric can be expressed most generally as
\cite{Bardeen,Misao,Mukhanov}
\begin{equation}
\label{metric}
ds^2 = (1+2A) dt^2 - 2\partial_i B \,dx^i dt
	- a(t)^2 \left[ (1+2{\cal R}) \delta_{ij}
	+ 2 \partial_i \partial_j E \right] dx^i dx^j \,.
\end{equation}
We follow \cite{Misao} most closely.
${\cal R}$ is the intrinsic curvature perturbation of the constant time
hypersurfaces.
On comoving hypersurfaces, $ \dot{\cal R}_c = HA_c $.
On flat hypersurfaces, $ {\cal R}_f \equiv 0 $.

\subsection{${\cal N}$}

Let $\{\Sigma(t)\}$ be a foliation of spacetime with hypersurfaces
$\Sigma(t)$ labeled by a certain coordinate time $t$ and let
$v^\mu$ be the unit vector field normal to $\Sigma(t)$.
Then $\theta = {v^\mu}_{;\mu} $ is the volume expansion rate of the
hypersurfaces along the integral curve $\gamma(\tau)$ of $v^\mu$.
For each integral curve, define
\begin{equation}
{\cal N} = \int_{\gamma(\tau)}\frac{1}{3}\theta d\tau \,,
\end{equation}
where $\tau$ is the proper time along the curve.

\subsection{${\cal R}$ and ${\cal N}$}

{}From \cite{Misao}
\begin{equation}
\label{theta}
\frac{1}{3} \theta = H \left( 1 - A + \frac{1}{H} \dot{\cal R}
	+\frac{1}{3H} \frac{1}{a^2}\partial^i\partial_i S_g\right) \,,
\end{equation}
where $ S_g = a^2 \dot{E} - B $. (If one Fourier expands $S_g$,
$ S_g = \sigma_g / q $, where $q=k/a$, in the notation of \cite{Misao}.)
Assuming that the anisotropic stress perturbation is negligible, which is
the case for scalar field, radiation or dust perturbations, then
the spatial trace-free part of the Einstein equations gives
\begin{equation}
\frac{1}{H} \dot{S}_g + S_g = \frac{1}{H} ( A + {\cal R} ) \,.
\end{equation}
{}From this equation we see that $S_g$ is at most of order $A/H$ or
${\cal R}/H$ so that it is clear that the last term in Eq.~(\ref{theta}) is
negligible compared with the other terms on superhorizon scales, that is for
$ q^2 \ll H^2 $ when the perturbations are Fourier expanded.
 From now on we work on superhorizon scales, and so we get
\begin{equation}
\frac{1}{3} \theta \simeq H \left( 1 - A + \frac{1}{H} \dot{\cal R} \right) \,.
\end{equation}
Also, from Eq.~(\ref{metric}),
\begin{equation}
d\tau = \left( 1 + A \right) dt \,.
\end{equation}
Therefore
\begin{equation}
{\cal N} = \int_{\gamma(\tau)} H \left( 1 - A + \frac{1}{H} \dot{\cal R}
	\right) \left( 1 + A \right) dt
 = \int_{\gamma(\tau)} \left( H + \dot{\cal R} \right) dt \,,
\end{equation}
and so
\begin{equation}
\delta N \equiv {\cal N} - N = \Delta {\cal R} \,.
\end{equation}
In particular, if we choose a foliation such that the initial hypersurface
is flat and the final one is comoving, we get
\begin{equation}
\delta N(\Sigma_f(t_1),\Sigma_c(t_2);\gamma(\tau)) = {\cal R}_c(t_2)
\end{equation}
for a given curve $\gamma(\tau)$.

Now take $t_1$ to be some time during inflation soon after the relevant scale
has passed outside the horizon and $t_2$ to be some time after complete
reheating\footnote{We do not consider the case of isocurvature
perturbations that persist until the present.}
when ${\cal R}_c$ has become constant.
The relevant scale is assumed to be still well outside the horizon at $t=t_2$.
Then one may regard ${\cal N}$ as a function of the field configuration
$\phi^a(t_1,x^i)$ on $\Sigma(t_1)$ and the time $t_2$,
\begin{equation}
{\cal N} = {\cal N} \left( \phi^a(t_1,x^i) , t_2 \right) \,.
\end{equation}
Note that in general ${\cal N}$ depends on both $\phi^a(t_1)$ and
$\dot{\phi}^a(t_1)$, but as $t_1$ is during inflation we use the slow roll
approximation to eliminate the dependence on $\dot{\phi}^a(t_1)$.
Therefore
\begin{equation}
\label{Rphi}
{\cal R}_c (t_2,x^i_2) = \delta N
= \frac{\partial N}{\partial \phi^a} \delta\phi^{a}_{f}(t_1,x^i_1) \,,
\end{equation}
where $x^i_1$ and $x^i_2$ are the spatial coordinates of $\gamma(\tau)$ on
$\Sigma_f(t_1)$ and $\Sigma_c(t_2)$, respectively.
One can of course choose the spatial coordinates on the hypersurfaces
by the condition $B=0$ to make $x^i_1=x^i_2$.
Since the perturbations in both $\theta$ and the density are negligibly small
on comoving hypersurfaces on superhorizon scales, $\Sigma_c(t_2)$ may
be regarded as a hypersurface of constant Hubble parameter or constant energy
density.

\section{Scalar fields}

We assume the scalar field part of the action to be
\begin{equation}
S = \int \left[
 \frac{1}{2} h_{ab} g^{\mu\nu} \partial_\mu \phi^a \partial_\nu \phi^b
 - V(\phi) \right] \sqrt{-g}\, d^4x \,.
\end{equation}
where $g_{\mu\nu}$ is the spacetime metric and $h_{ab}$ is the metric on the
scalar field space.

\subsection{The background}

The background scalar fields are spatially homogeneous
\begin{equation}
\phi^a = \phi^a (t) \,.
\end{equation}
The following formula will be useful
\begin{equation}
\label{Nphi}
\frac{\partial N}{\partial \phi^a} \dot{\phi}^a = -H \,.
\end{equation}
The background equation of motion for the scalar fields is
\begin{equation}
\frac{D\dot{\phi}^a}{dt} + 3H \dot{\phi}^a + h^{ab} V_{,b} = 0 \,,
\end{equation}
where $ DX^a = dX^a + {\Gamma^{a}}_{bc} X^b d\phi^c $ and
$ {\Gamma^{a}}_{bc} = \frac{1}{2} h^{ad}
	\left( h_{db,c} + h_{dc,b} - h_{bc,d} \right) $.
We assume that the scalar potential is sufficiently flat, \mbox{i.e.}
satisfies
\begin{equation}
\label{sr1}
V^{;a} V_{;a} \ll V^2
\hspace{1cm}{\rm and}\hspace{1cm}
\sqrt{ V^{;ab} V_{;ab} } \ll V \,,
\end{equation}
where the semicolon denotes the covariant derivative in the scalar field
space.
Then the scalar field dynamics will rapidly approach slow roll given by
\begin{equation}
3H \dot{\phi}^a \simeq - h^{ab} V_{,b} \,,
\end{equation}
or
\begin{equation}
\label{bsr}
\frac{\dot{\phi}^a}{H} \simeq - \frac{V^{;a}}{V} \,.
\end{equation}
We assume that slow roll has been attained for all epochs of interest.

\subsection{$\delta\phi^a$}

The equation of motion for the Fourier modes of scalar field
perturbations on flat hypersurfaces is
\begin{equation}
\frac{ D^2 \delta\phi^{a}_{\bf k} }{ dt^2 }
 + 3H \frac{ D \delta\phi^{a}_{\bf k} }{ dt }
 - {R^a}_{bcd} \dot{\phi}^b \dot{\phi}^c \delta\phi^{d}_{\bf k}
 + q^2 \delta\phi^{a}_{\bf k} + V^{;a}{}_{;b} \delta\phi^{b}_{\bf k}
= \frac{1}{a^3} \frac{D}{dt} \left( \frac{a^3}{H} \dot{\phi}^a
	\dot{\phi}^b \right) h_{bc} \delta\phi^{c}_{\bf k}
\end{equation}
where ${\bf k}$ is the comoving wavenumber, $ q = |{\bf k}| / a $,
and the scalar field space curvature
$ {R^a}_{bcd} = {\Gamma^{a}}_{bd,c} - {\Gamma^{a}}_{bc,d}
 + {\Gamma^{a}}_{ce} {\Gamma^{e}}_{db}
 - {\Gamma^{a}}_{de} {\Gamma^{e}}_{cb} $.
Let $\Delta\phi$ denote the characteristic curvature radius of the
scalar field space.
We assume $ h_{ab} \dot{\phi}^a \dot{\phi}^b / (\Delta\phi)^2 \ll H^2 $ or
\begin{equation}
\label{sr2}
\frac{ V^{;a} V_{;a} }{ V^2 } \ll (\Delta\phi)^2 \,.
\end{equation}
This is automatic if $\Delta\phi \sim 1$ as is typically the case.
Then modes well outside the horizon, \mbox{i.e.} with $ q^2 \ll H^2 $,
satisfy the slow roll equation of motion
\begin{equation}
\label{sr}
3H \frac{D \delta\phi^{a}_{\bf k} }{dt}
 - {R^a}_{bcd} \dot{\phi}^b \dot{\phi}^c \delta\phi^{d}_{\bf k}
 + V^{;a}{}_{;b} \delta\phi^{b}_{\bf k}
 = 3 \dot{\phi}^a \dot{\phi}^b h_{bc} \delta\phi^{c}_{\bf k} \,.
\end{equation}

\section{Calculation of the spectral index}

Fourier expansion of Eq.~(\ref{Rphi}) gives
\begin{equation}
{\cal R}_{\bf k}(t_2)
 = \frac{\partial N}{\partial \phi^a} \delta\phi_{\bf k}^a(t_1) \,.
\end{equation}
For scalar field perturbations generated from vacuum fluctuations during
inflation
\begin{equation}
\delta\phi_{\bf k}^a
 = \sum_\alpha \phi_{k}^{a\alpha} a_{\bf k}^{\alpha} \,,
\end{equation}
where $a_{\bf k}^{\alpha}$ is a
classical\footnote{We are outside the horizon so everything is classical.}
random variable satisfying
\begin{equation}
\langle a_{\bf k}^{\alpha} {a_{\bf l}^\beta}^{\dagger} \rangle
 = \delta^{\alpha\beta} \delta^3 ({\bf k-l}) \,,
\end{equation}
$\alpha$ runs over the number of scalar field components, and
$\phi_{k}^{a\alpha}$ is real and satisfies
\begin{equation}
\label{norm}
\sum_\alpha \phi_{k}^{a\alpha} \phi_{k}^{b\alpha}
= \frac{H^2}{2k^3} \left( h^{ab} + \epsilon^{ab} \right)
\end{equation}
where, assuming the slow roll conditions Eqs.~(\ref{sr1})
and~(\ref{sr2}), $\epsilon^{ab}$ is small and slowly varying with respect to
$a$ at fixed $k/a$.
Therefore the spectrum of curvature perturbations is given by
\begin{eqnarray}
\frac{2\pi^2}{k^3} P_{\cal R} \delta^3 ({\bf k-l})
 & = &
\langle {\cal R}_{\bf k}(t_2) {\cal R}_{\bf l}^{\dagger}(t_2) \rangle \\
 & = &
\frac{\partial N}{\partial \phi^a} \frac{\partial N}{\partial \phi^b}
\langle \delta\phi_{\bf k}^a(t_1)
	{\delta\phi_{\bf l}^b}^{\dagger}(t_1) \rangle \\
 & = & \sum_{\alpha}
\frac{\partial N}{\partial \phi^a} \frac{\partial N}{\partial \phi^b}
\phi_{k}^{a\alpha} \phi_{k}^{b\alpha} \delta^3 ({\bf k-l}) \,,
\end{eqnarray}
or
\begin{equation}
\label{P}
P_{\cal R} = \frac{k^3}{2\pi^2} \sum_{\alpha}
\frac{\partial N}{\partial \phi^a} \frac{\partial N}{\partial \phi^b}
\phi_{k}^{a\alpha} \phi_{k}^{b\alpha} \,.
\end{equation}
The spectral index is then given by
\begin{equation}
\label{n}
n_{\cal R} - 1 = \frac{d \ln P_{\cal R}}{d \ln k}
= 3 + \frac{k^3}{2\pi^2 P_{\cal R}} \sum_{\alpha}
\frac{\partial N}{\partial \phi^a} \frac{\partial N}{\partial \phi^b}
\frac{D\left(\phi_{k}^{a\alpha}\phi_{k}^{b\alpha}\right)}{\partial \ln k} \,.
\end{equation}
Note that $ \sum_{\alpha} \phi_{k}^{a\alpha} \phi_{k}^{b\alpha} $,
which is given by Eq.~(\ref{norm}), is to be evaluated at a fixed time
$t_1$. The $k$-dependence of the spectrum (apart from the trivial $k^3$
factor) is hidden in the small term $\epsilon^{ab}$.
We will exploit the fact that $\epsilon^{ab}$ is small and slowly varying
with respect to $a$ at fixed $k/a$ in order to evaluate it as follows.
\begin{equation}
\left. \frac{D}{\partial \ln k} \right|_{a={\rm constant}}
= \left. \frac{D}{\partial \ln a} \right|_{k/a={\rm constant}}
- \left. \frac{D}{\partial \ln a}\right|_{k={\rm constant}} \,,
\end{equation}
Eq.~(\ref{norm}) gives
\begin{equation}
\left. \frac{D}{\partial \ln a} \left( \sum_\alpha \phi_{k}^{a\alpha}
	\phi_{k}^{b\alpha} \right) \right|_{k/a={\rm constant}}
= - \left( 3 - 2 \frac{\dot{H}}{H^2} \right)
	\sum_\alpha \phi_{k}^{a\alpha} \phi_{k}^{b\alpha} \,,
\end{equation}
and the slow roll equation of motion Eq.~(\ref{sr}) gives
\begin{equation}
\left. \frac{D \phi_{k}^{a\alpha}}{\partial \ln a}
\right|_{k={\rm constant}}
= \left( \frac{\dot{\phi}^a \dot{\phi}^b}{H^2} h_{bd}
 + \frac{1}{3} {R^a}_{bcd} \frac{\dot{\phi}^b \dot{\phi}^c}{H^2}
 - \frac{V^{;a}{}_{;d}}{V} \right) \phi_{k}^{d\alpha} \,.
\end{equation}
Substituting into Eq.~(\ref{n}) then gives
\begin{equation}
\label{n2}
n_{\cal R} - 1 = 2 \frac{\dot{H}}{H^2}
- 2 \frac{k^3}{2\pi^2 P_{\cal R}} \sum_{\alpha}
\frac{\partial N}{\partial \phi^a}
\left( \frac{\dot{\phi}^a \dot{\phi}^b}{H^2} h_{bd}
 + \frac{1}{3} {R^a}_{bcd} \frac{\dot{\phi}^b \dot{\phi}^c}{H^2}
 - \frac{V^{;a}{}_{;d}}{V} \right) \phi_{k}^{d\alpha}
\phi_{k}^{e\alpha} \frac{\partial N}{\partial \phi^e} \,.
\end{equation}
Now from Eq.~(\ref{norm})
\begin{equation}
\sum_\alpha \phi_{k}^{a\alpha} \phi_{k}^{b\alpha}
\simeq \frac{H^2}{2k^3} h^{ab} \,,
\end{equation}
and so from Eq.~(\ref{P})
\begin{equation}
P_{\cal R} = \left( \frac{H}{2\pi} \right)^2 h^{ab}
\frac{\partial N}{\partial \phi^a} \frac{\partial N}{\partial \phi^b} \,,
\end{equation}
therefore from Eq.~(\ref{n2})
\begin{equation}
n_{\cal R} - 1 = 2 \frac{\dot{H}}{H^2}
- 2 \frac{
 \frac{\partial N}{\partial \phi^a}
 \left( \frac{\dot{\phi}^a \dot{\phi}^d}{H^2}
 + \frac{1}{3} {{R^a}_{bc}}^{d} \frac{\dot{\phi}^b \dot{\phi}^c}{H^2}
 - \frac{V^{;ad}}{V} \right)
 \frac{\partial N}{\partial \phi^d} }
{ h^{ef} \frac{\partial N}{\partial \phi^e}
	\frac{\partial N}{\partial \phi^f} } \,.
\end{equation}
Therefore, from Eqs.~(\ref{Nphi}) and~(\ref{bsr}),
\begin{eqnarray}
n_{\cal R} - 1 & = & 2 \frac{\dot{H}}{H^2}
- 2 \frac{ 1 + \frac{\partial N}{\partial \phi^a}
 \left( \frac{1}{3} R^{abcd} \frac{V_{;b}V_{;c}}{V^2}
  - \frac{V^{;ad}}{V} \right)
 \frac{\partial N}{\partial \phi^d} }
{ h^{ef} \frac{\partial N}{\partial \phi^e}
 \frac{\partial N}{\partial \phi^f} } \,, \\
& = & \frac{ \left[ 2 {(\ln V)^{;a}}_{;b}
 + \left( \frac{2}{3} {R^{a}}_{cbd} - {h^{a}}_{b} h_{cd} \right)
 (\ln V)^{;c} (\ln V)^{;d} \right] N_{;a} N^{;b} }{ N_{;e} N^{;e} } \,.
\end{eqnarray}

\section{Summary}

The spectrum of gravitational waves produced during inflation is \cite{grav}
\begin{equation}
P_g = \left( \frac{H}{2\pi} \right)^2 \,,
\end{equation}
where, here and below, the right hand side of the equation is
to be evaluated at the time when the relevant scale passed through the
Hubble radius during inflation, \mbox{i.e.} at $aH=k$.
The spectrum of curvature perturbations produced during inflation is
\begin{equation}
P_{\cal R} = \left( \frac{H}{2\pi} \right)^2
h^{ab} \frac{\partial N}{\partial \phi^a} \frac{\partial N}{\partial \phi^b}
\,,
\end{equation}
where $N$ is the number of $e$-folds of expansion from a given point
in scalar field space during inflation to some reference energy
density or Hubble parameter during radiation domination, \mbox{i.e.}
after complete reheating.\footnote{We do not consider the case of
isocurvature perturbations that persist until the present.}
The spectral index of the gravitational waves is \cite{index}
\begin{equation}
n_g = 2 \frac{\dot{H}}{H^2} \,.
\end{equation}
The spectral index of the curvature perturbations is
\begin{eqnarray}
n_{\cal R} - 1 & = & 2 \frac{\dot{H}}{H^2}
- 2 \frac{ 1 + \frac{\partial N}{\partial \phi^a}
 \left( \frac{1}{3} R^{abcd} \frac{V_{;b}V_{;c}}{V^2}
  - \frac{V^{;ad}}{V} \right)
 \frac{\partial N}{\partial \phi^d} }
{ h^{ef} \frac{\partial N}{\partial \phi^e}
 \frac{\partial N}{\partial \phi^f} } \,, \\
& = & \frac{ \left[ 2 {(\ln V)^{;a}}_{;b}
 + \left( \frac{2}{3} {R^{a}}_{cbd} - {h^{a}}_{b} h_{cd} \right)
 (\ln V)^{;c} (\ln V)^{;d} \right] N_{;a} N^{;b} }{ N_{;e} N^{;e} } \,.
\end{eqnarray}
Using Eq.~(\ref{Nphi}) we see that inflation predicts
\begin{equation}
\label{constraint}
\frac{P_g}{P_{\cal R}} \leq | n_g | \,,
\end{equation}
so that the ratio of gravitational wave to curvature perturbation
contributions to the cosmic microwave background anisotropy satisfies
$ R \lsim 5 | n_g | $.
\footnote{We follow \cite{Polarski} in using the factor of $\sim 5$.
The results of this paper, and in particular Eq.~(\ref{constraint}),
are not valid for an inflationary model that cannot be realized using any
number of interacting scalar fields minimally coupled to Einstein gravity
\cite{R2}.}
Note that the spectral index depends on the curvature of the space of
scalar fields as well as the potential, though as (in the opinion of EDS)
realistic models of inflation tend to give $ (V'/V)^2 \ll |V''/V| $
this may be difficult to observe.
An interesting point is that in models with more than one dynamical degree of
freedom there is generally not a unique inflationary trajectory and so the
initial conditions might play a role in determining the spectrum and hence be
observable.

\subsection*{Acknowledgements}
EDS thanks E. W. Kolb for stimulating comments and A. A. Starobinsky for a
stimulating seminar. We thank D. H. Lyth for helpful discussions and
D. H. Lyth, the referee and J. Garc\'{\i}a-Bellido for helpful
suggestions as to how to make this paper easier to follow.
The work of MS is supported by Monbusho Grant-in-Aid for Scientific Research
No.~05640342.
EDS was supported by the JSPS during the early stages of this work,
by the Royal Society when this work was completed, and is now
again supported by the JSPS.

\end{document}